\documentclass[10pt,a4paper]{article}

\usepackage{hyperref}
\usepackage{amsmath}
\usepackage{amssymb}
\usepackage{graphicx}

\usepackage{algorithmic}
\usepackage{algorithm}

\newcommand{\beq}[1][]{\begin{equation}\label{#1}}
\newcommand{\eeq}{\end{equation}}

\newcommand{\beqq}{\begin{equation*}}
\newcommand{\eeqq}{\end{equation*}}

\usepackage{cite}

\usepackage[bf,flushleft]{caption2}

\makeatletter
\renewcommand{\@biblabel}[1]{\quad#1.}
\makeatother

\date{}

\begin{document}
\begin{flushleft}
{\Large
\textbf{Multi-input  distributed classifiers for synthetic genetic circuits}
}
\\
Oleg Kanakov$^{1,\ast}$, Roman Kotelnikov$^{1}$, Ahmed
Alsaedi$^{2}$, Lev Tsimring$^{3}$, Ramon Huerta$^{3}$, Alexey
Zaikin$^{4,1}$, Mikhail Ivanchenko$^{1,\ast}$
\\
\bf{1}  Lobachevsky State University of Nizhniy Novgorod, Nizhniy
Novgorod, Russia
\\
\bf{2} Department of Mathematics, King AbdulAziz University,
Jeddah, Saudi Arabia
\\
\bf{3} BioCircuits Institute, University of California San Diego,
La Jolla CA, USA
\\
\bf{4} Institute for Women's Health and Department of Mathematics,
University College London, London, United Kingdom
\\
$\ast$ E-mail: okanakov@rf.unn.ru
\end{flushleft}

\section*{Abstract}

For practical construction of complex synthetic genetic networks able to perform elaborate functions
 it is important to have a pool of relatively simple ``bio-bricks'' with different functionality which can be compounded together.
To complement engineering of very different existing synthetic genetic devices such as switches, oscillators or logical gates,
we propose and develop here a design  of synthetic multiple input distributed classifier with learning ability.
Proposed classifier will be able to separate multi-input data, which are inseparable for single input classifiers. Additionally, the data classes could potentially occupy the area of any shape in the space of inputs. We study two approaches to classification, including hard and soft classification and confirm the schemes of genetic networks by analytical and numerical results.


\section*{Introduction}

The current challenge facing the synthetic biology research
community is the construction of relatively simple, robust and
reliable genetic networks, which will mount a pool of
``bio-bricks'', potentially to be connected into more complex
systems. Rapid progress of experimental synthetic biology has
indeed provided several synthetic genetic networks with different
functionality. Since the year $2000$ with the development of two fundamental
simple networks, representing the toggle switch
\cite{2000_Gardner} and the repressilator \cite{2000_Elowitz},
there have been a vast number of proof-of-principle synthetic
networks designed and engineered. To enumerate some of them,
functionally these circuits included transcriptional or metabolic
oscillators \cite{2008_Stricker,2009_Tigges,2005_Fung}, spatially
coupled and synchronised oscillators \cite{2010_Danino,2011_Kim},
calculators \cite{2009_Friedland}, inducers of pattern formation
\cite{2005_Basu}, learning systems \cite{2009_Fernando},
optogenetic devices \cite{2005_Levskaya}, memory circuits and
logic gates \cite{2012_Bonnet,2011_Tamsir,2013_Bonnet,2013_Siuti}.

One of the much awaited kinds of synthetic gene circuits with
principally new functionality would work as intelligent
biosensors, for example, realized as genetic classifiers able to
assign inputs with different classes of outputs. Importantly, they
would need to allow an arbitrary shape of the area in the space of
inputs, in contrast to simple threshold devices. Recently, the
first step in this direction has been made in \cite{FirstPaper},
where the concept of a distributed genetic classifier formed by a
population of genetically engineered cells has been proposed. Each
cell constituting the distributed classifier is essentially an
individual binary classifier with specific parameters, which are
randomly varied among the cells in the population. The inputs to
the classifier are certain chemical concentrations, which the
engineered cells can be made sensitive to. The classification
decision can be read out from an individual cell, for example, by
the fluorescent protein technique which is well developed and
universally adopted in synthetic biology. The output of the whole
distributed classifier is the sum of the individual classifier
outputs, and the overall decision is made by comparing this output
to a preset threshold value. If the initial (or ``master'')
population contains a sufficiently diverse variety of cells with
different parameters, the whole ensemble can be trained by
examples to solve a specific classification problem just by
eliminating the cells which answer incorrectly to the examples
from the training sequence, without tuning any parameters of the
individual classifiers.

The paper \cite{FirstPaper} focused on distributed classifiers
composed of single-input elementary classifiers. The single-input
genetic circuit proposed in \cite{FirstPaper} provides a
bell-shaped output function against the input chemical
concentration. The individual cells in the population differ from
each other by the choice of the particular input chemicals that
they are sensitive to, and the width and positioning of the
bell-shaped response function. These parameters can be varied in a
range of up to $10^5$ by modifying the ribosome binding sites in
the gene circuit \cite{salis2009automated,kudla2009coding}. Such
libraries of cells with randomized individual parameters have been
constructed in experiments for synthetic circuit optimization
\cite{pfleger2006combinatorial,wang2009programming,zelcbuch2013spanning}
The single-input distributed classifier has been tested on several
examples in \cite{FirstPaper}.

However, practical applications may require classification of
multiple inputs. In \cite{FirstPaper} it has been discussed that
the same principles can be utilized for a design of two- or
multi-input circuits. The proposed circuit is based upon a genetic
AND gate \cite{wang2011engineering,moon2012genetic,shis2013library},
providing a bell-shaped response function in the space of two or
more inputs. Nevertheless, no studies of a distributed classifier
with two or more inputs have been performed so far. In this paper
we fill this gap by developing distributed classifiers based upon
two types of elementary two-input classifier cells: one is a
simple scheme implementing a linear classifier in the space of two
inputs and the other is the scheme with AND gate and bell-shaped
response proposed in \cite{FirstPaper}.

Following this we consider two settings of the classification
problem. In the first setting, which we refer to as ``hard
classification'', the classes are assumed separable, which implies
that the sets of points belonging to either class in the parameter
space do not intersect. In this case all elementary classifiers
can be unambiguously separated into those answering correctly and
incorrectly to the training examples, and the ``hard learning
strategy'' may be used, which is based upon discarding all
incorrectly answering cells.

We start with considering the case of separable classes and hard
learning, using linear classifiers as elementary cells. We show,
that a range of separable classification problems can be reliably
solved even with a little number of elementary classifiers using
this strategy, including problems which become inseparable (and,
thus, imposing a lower bound on the error rate, which can not be
subdued) when attempted to be solved by single-input classifiers.
At the same time, this approach is incapable of solving
classification problems with more complicated classification
borders, as well as problems with inseparable classes.

In the second part of our paper we address both mentioned
issues by means of soft learning strategy and elementary cells
with bell-shaped response. We demonstrate the effectiveness of
this approach for solving these more complicated tasks at the
expense of a more complicated gene circuit in each elementary
classifier and a greater number of cells required.

\section*{Hard classification problem}\label{sec_hard}


\subsection*{Two-input linear classifier circuit}
\begin{figure}
\centering
\includegraphics[width=0.8\textwidth,clip=true]{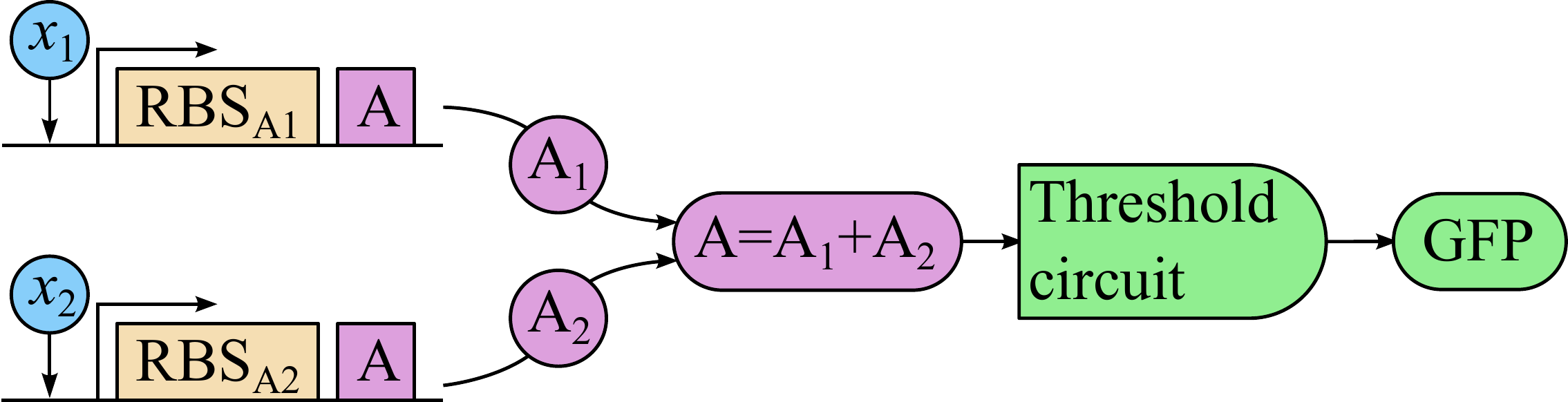}
\caption{{\bfseries Scheme of a two-input linear classifier
circuit.} $x_1$, $x_2$ -- inputs inducing the corresponding
promoters, $\text{RBS}_{\text{A1}}$ and $\text{RBS}_{\text{A2}}$
-- ribosome binding sites determining the strengths of the input
branches, $\text{A}$ -- intermediate transcription factor (same in
both input branches), GFP -- reporter gene.}\label{fig_lincircuit}
\end{figure}

We assume, that the classifier input is a set of chemical
concentrations capable of regulating appropriate synthetic
promoters (directly or mediated by the regulatory network of the
cell). In the simplest design of a multi-input genetic classifier
circuit, the input genes drive the synthesis of the same
intermediate transcription factor $A$ (Fig.~\ref{fig_lincircuit}),
but are regulated by different promoters sensitive to the
corresponding input chemicals $X_{j}$. The expression of the
reporter protein, for example, green fluorescent protein (GFP), is
driven by the total concentration of $A$, summarized from all
input genes.

The stationary concentration $a$ of the intermediate transcription
factor can be expressed as a weighted sum over all classifier
inputs
\beq
a=\sum_j b_j a_j(x_j),
\eeq
where $x_j$ are concentrations of the inputs $X_j$, $a_j(\cdot)$
are nonlinear functions, each describing the response to a
particular input, including the whole appropriate signalling
pathway, and $b_j$ are linear multipliers determining the relative
strengths of the corresponding inputs, which can be varied in a
range of more than $10^5$ fold by varying the DNA sequence within
and near the ribosome binding site of the corresponding input gene
\cite{salis2009automated,kudla2009coding}.

For a sharper discrimination between the classifier decisions, we
propose to make use of the protein sequestration technique
\cite{buchler2009protein} to generate an ultrasensitive response
to $A$ when its concentration exceeds a certain threshold. This is
achieved by binding $A$, which normally induces the reporter gene,
with a suitable inhibitor into an inactive complex which can not
bind DNA. The simplest description of this binding assumes that
free active transcription factor $A$ becomes available only when
all inhibitor molecules are bound. Then the reporter protein
concentration $g$ may be approximated by a shifted and truncated
Hill function \cite{buchler2009protein}
\beq\label{eq_sequestr}
g=g(a;\theta)=\begin{cases}
 \alpha \gamma, & \mbox{if}\, a \le \theta,\\
 \gamma \frac{\alpha A_g + a-\theta}{A_g+a-\theta}, & \mbox{if}\, a > \theta,
\end{cases}
\eeq
where $\theta$ is the threshold determined by the constitutive
expression rate of the inhibitor \cite{buchler2009protein}, $A_g$
is the DNA-binding dissociation constant for $A$, $\gamma$
determines the maximal output, and $\alpha \gamma$ is the basal
expression of the reporter protein in the absence of $A$.

A master population of cells with randomized individual response
characteristics can be obtained by randomly varying the input
weights $b_j$, as well as the threshold $\theta$, among the cells
in the population. In the following we restrict ourselves to the
case of two inputs, but our approach equally applies to input
vectors of any dimension. We assume, that the parameter values in
the $i$th individual cell are $b_1^i$ and $b_2^i$ for the input
weights and $\theta^i$ for the threshold, the lower index denoting
the input and the upper one labeling the cells, all other
parameters being the same in both input channels in all cells.

The GFP output of a chosen $i$th individual classifier cell is
then
\beq
f_i(x_1,x_2)=g(b_1^i a_1(x_1) + b_2^i a_2(x_2); \theta^i)
\eeq
with $g(a;\theta)$ defined in \eqref{eq_sequestr}.

We use the discrete-output model of the individual cell to analyze
the learning process and the distributed classifier behaviour.
Namely, we assume, that each individual cell can produce two
distinguishable kinds of output, corresponding to the cases in
\eqref{eq_sequestr}: low, or ``negative answer'' (which is the
subthreshold background output $g_i=\alpha\gamma$), and high, or
``positive answer'' (above-threshold output).

We note, that each individual cell acts as a linear classifier in
the transformed input space with coordinates $(a_1, a_2)$ defined
by the corresponding nonlinear input functions
\beq
a_1=a_1(x_1), \quad a_2=a_2(x_2).
\eeq
Indeed, an individual $i$th cell generates high output, when
$b^i_1 a_1 + b^i_2 a_2>\theta^i$, or
\beq\label{eq_pos_response}
m_1^i a_1 + m_2^i a_2 > 1,
\eeq
where $m_{1,2}^i=b_{1,2}^i/\theta^i$.

Such classifier divides the transformed input space into two
regions, corresponding to either answer of the classifier, which
we will refer to as the negative and the positive classes. The
border separating the classes in the transformed input space is a
straight line
\beq\label{eq_Aborder}
m^i_1 a_1 + m^i_2 a_2 =1.
\eeq
Note, that $a_{1,2}$ as well as $m_{1,2}$ can not be negative due
to their meaning. In the following, $a_{1,2}$ and $m_{1,2}$ are
assumed to be non-negative real numbers. In particular it means,
that the space of inputs and the space of parameters are always
limited to the first quadrant of the full real space, regardless
of its dimension.

\subsection*{Hard classification principle and learning
strategy}\label{sec_hard_principle}

\begin{figure}
\centering
\includegraphics[width=0.4\textwidth,clip=true]{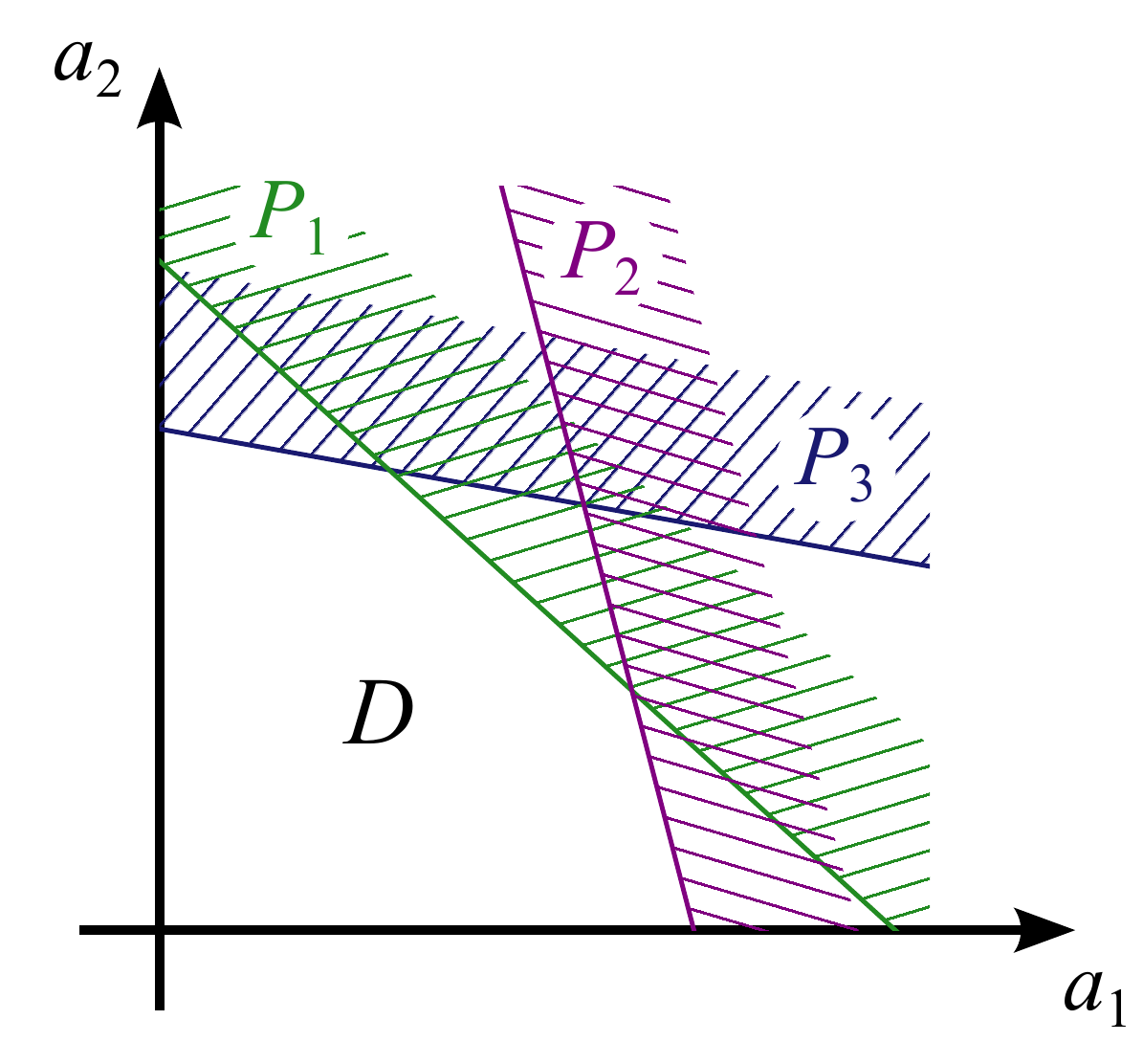}
\caption{{\bfseries Hard classification principle.} $P_1$, $P_2$,
$P_3$ -- positive classes of individual linear classifiers, $D$ --
negative class of the collective classifier.}\label{fig_cons3}
\end{figure}

An ensemble of linear classifiers can be utilized to perform a
more complicated classification task with a piecewise-linear
border in the transformed input space. Denote with $P_i$ the
positive class of the $i$th individual classifier:
\beq\label{eq_Aplus}
P_i=\left\{a_1, a_2: m_1^i a_1 + m_2^i a_2 > 1\right\}.
\eeq
Let all elements in the ensemble be given the same input. Then the
whole ensemble can be used as a single distributed classifier,
dividing the transformed input space into the positive class
$P=\bigcup_i P_i$, where at least one individual classifier gives
the positive answer, and the negative class $D=\bar P=\bigcap_i
\bar P_i$, where all classifiers answer negatively (here the bar
``$\bar\;$'' denotes complement in the transformed input space),
see Fig.~\ref{fig_cons3}.

By construction, the negative class $D$ is entirely contained in
each closed half-plane defined by any of its edges, which means it
is always convex. The classification border is a polygonal line
composed of segments, each described by an equation of type
\eqref{eq_Aborder}, all having negative slope, because both
$m^i_1$ and $m^i_2$ are positive. In the limit of large number of
cells, the negative class becomes a convex region bordered by the
coordinate axes and a smooth classification border having negative
tangent slope at each point.

An ensemble constituting a distributed classifier with a specified
(``target'') classification border (satisfying the requirements of
negative slopes and convexity) can be prepared by the following
learning algorithm. Let us start with a master population of
linear classifiers of type \eqref{eq_Aplus} with random parameters
$m_1^i$, $m_2^i$ distributed continuously over some interval. The
aim of the learning is to keep all individual classifiers which
answer correctly to all training examples and remove all
incorrectly answering ones. To achieve this, we test the whole
ensemble against a training sequence of samples from the negative
class. All elements which answer positively to at least one
negative sample are considered ``incorrect'' and are removed from
the ensemble. This can be done, for example, using the
fluorescence-activated cell sorting (FACS) technique. Positive
class samples are not needed for learning, since hard
classification fundamentally assumes separability of classes.

Actually, it is enough to use only samples located along the
classification border. Although training sequences of this kind
might be not available in real situations, theoretically,
excluding the interior of the negative region from the training
sequence leads to achieving the same learning outcome with a
smaller number of samples.

The ensemble which remains after this learning procedure forms a
distributed classifier with the class border determined by the
training sequence. The actual set of cells constituting the
trained distributed classifier is essentially the outcome of
clipping the master population in the parameter space $(m_1, m_2)$
with a certain mask, which completely characterizes the action of
the learning algorithm. In other words, the trained ensemble is a
set intersection of the master population with a region in the
parameter space, which we will refer to as the ``trained ensemble
region''.

\begin{figure}
\centering
(a)\includegraphics[width=0.4\textwidth,clip=true]{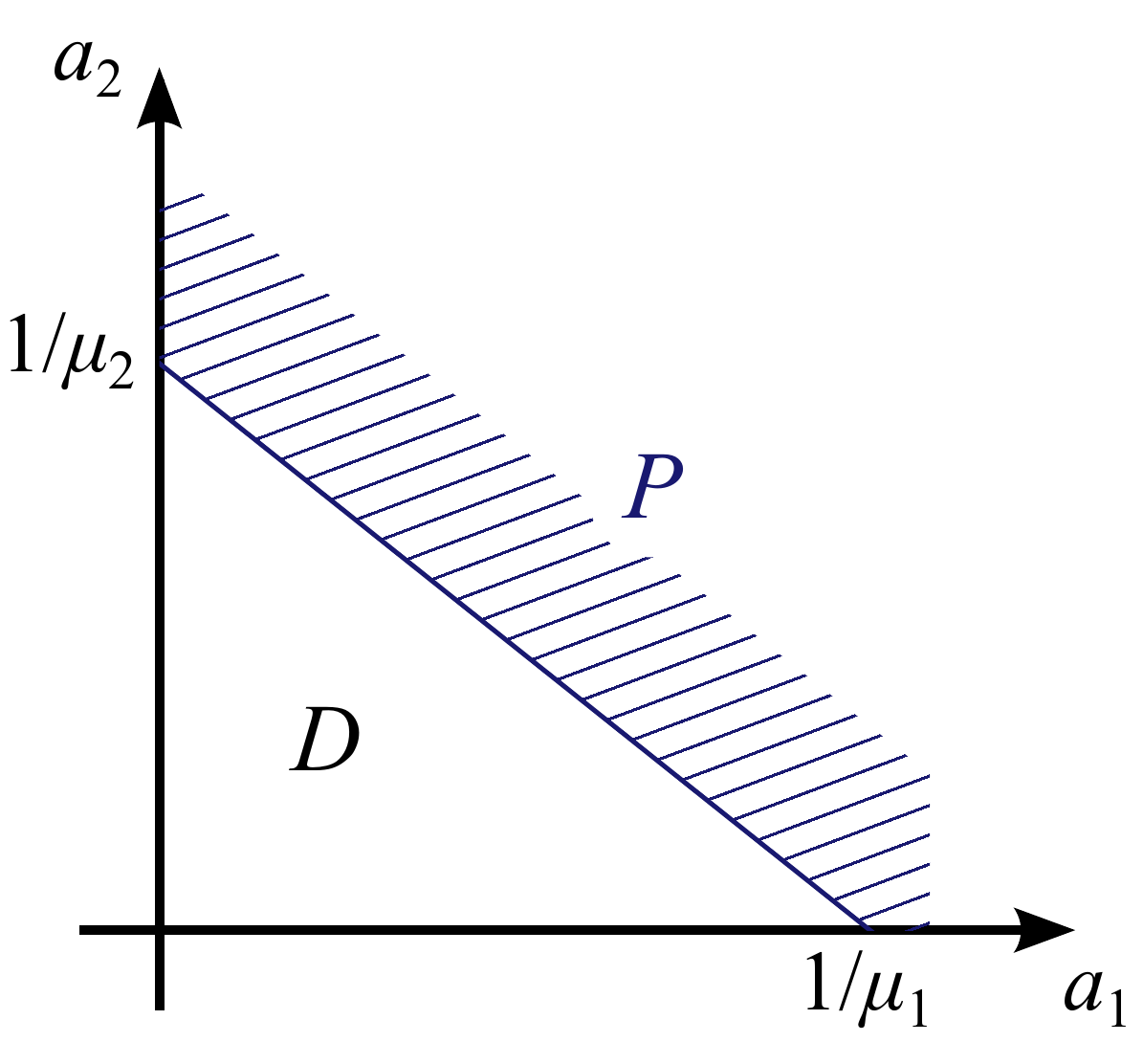}
(b)\includegraphics[width=0.4\textwidth,clip=true]{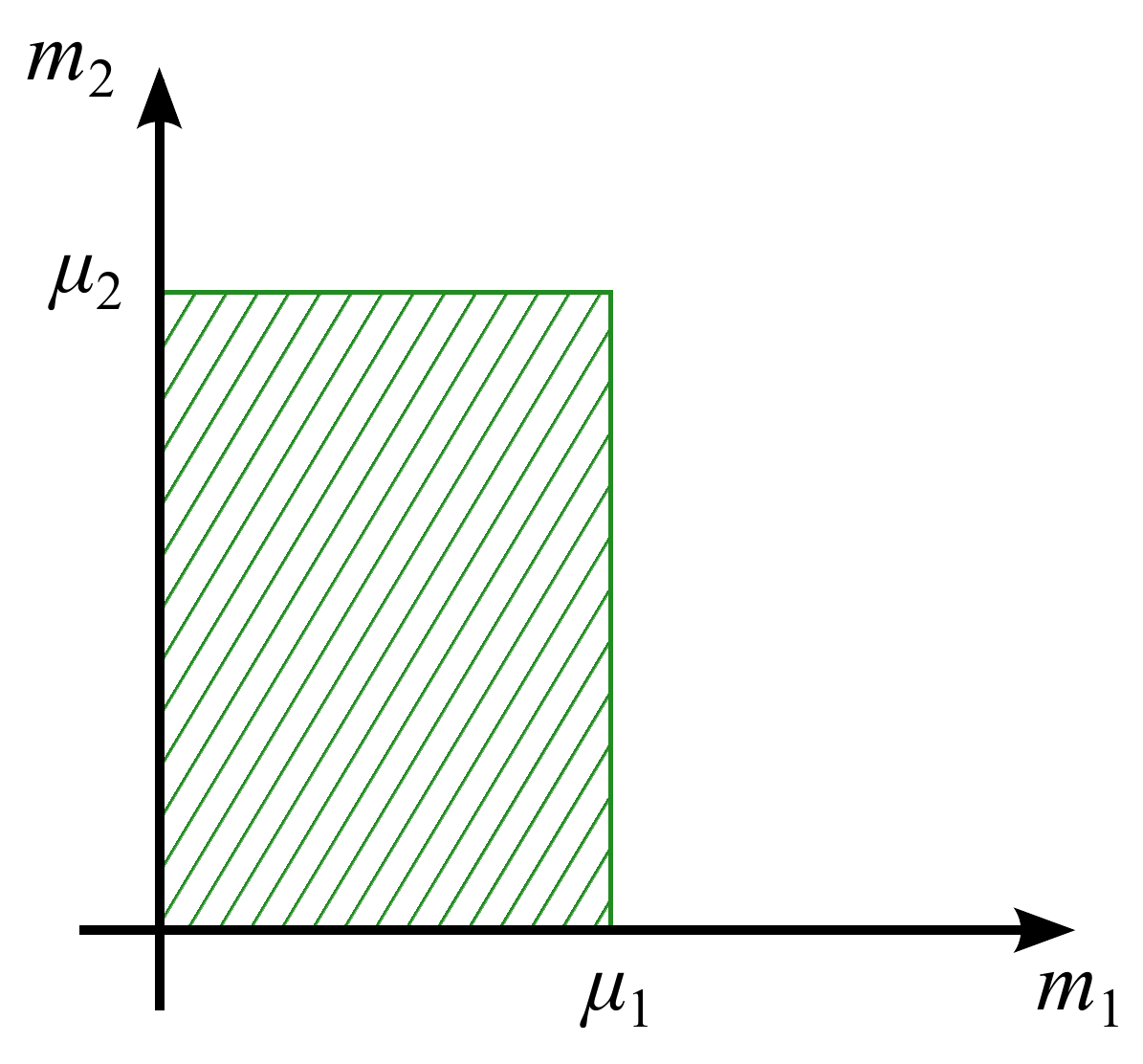}
\caption{{\bfseries Training a distributed classifier with a
linear target border.} {\itshape Panel (a).} Target classes: $P$
-- positive, $D$ -- negative. {\itshape Panel (b).} Trained
ensemble region on the plane of parameters: hatched
area.}\label{fig_muborder}
\end{figure}

To get an insight into a quantitative description of hard learning
strategy, we start with a trivial case when the target
classification border is linear, defined by the equation
\beq\label{eq_mubord}
\mu_1 a_1 + \mu_2 a_2 =1,
\eeq
where $\mu_{1,2}$ are given constant coefficients, see
Fig.~\ref{fig_muborder}~(a). Although this classification task can
be solved by a single linear classifier, we use it as a starting
point to describe the training of a distributed classifier.

In the course of learning with a sequence of points distributed
along the border \eqref{eq_mubord}, any element having $m_1>\mu_1$
or $m_2>\mu_2$ will eventually answer positively and therefore
will be removed from the ensemble. Thus, the trained ensemble
region on the plane $(m_1, m_2)$ is a rectangle (hatched area in
Fig.~\ref{fig_muborder}~(b)).

Similarly, if the target border is a polygonal line (satisfying
the requirements of negative slopes and convexity), with the
target positive class being a union of several linear classes
\beq\label{eq_polybord}
P=\bigcup_i \left\{a_1, a_2: \mu_1^{i} a_1 + \mu_2^{i} a_2 >1
\right\},
\eeq
where $\mu_1^{i}$, $\mu_2^{i}$ are the coefficients of the
individual segments of the target polygonal border, then the
trained ensemble region on the plane $(m_1, m_2)$ is a convex
polygon with vertices $(\mu_1^{i}, \mu_2^{i})$, shown in
Fig.~\ref{fig_polylearn} (a) in Appendix S1 (see Supporting
Information) as hatched area.

In Appendix S1 we analyze the response of a trained hard
classifier to an input taken from the positive class. In
particular, a lower estimate is obtained for the quantity of cells
answering positively to such inputs. It is found to be
proportional to the density of the master population per unit of
the logarithmic parameter space $(\log m_1, \log m_2)$. It is also
shown, that the maximal quantity $m_{\max}$, to which the region
covered by the master population in the parameter space extends in
both $m_1$ and $m_2$, should be not less than the inverse of the
smaller intercept of the target class border (the intercepts are
the abscissa and the ordinate of the points where the border
crosses the axes $Oa_1$ and $Oa_2$).

\subsection*{Simulations}
\begin{figure}
{\centering \includegraphics[width=0.8\textwidth]{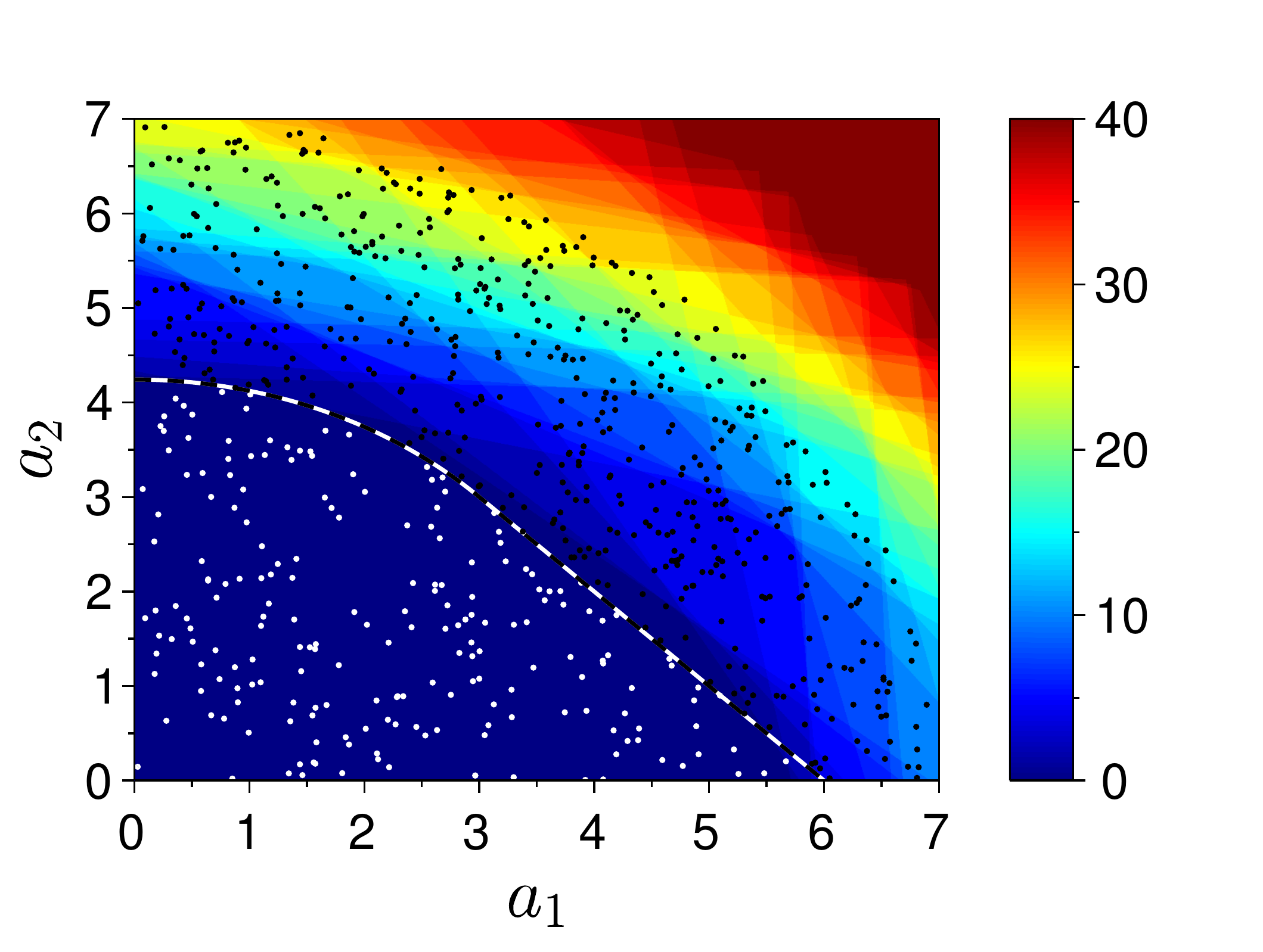}
\caption{{\bfseries Simulation results for hard classification.}
Response of a trained distributed classifier in the space of
inputs. Black-white dashed line -- target (predefined) class
border, white (black) filled circles -- samples from the negative
(positive) class, color -- number of the positively responding
cells (quantities 40 and above marked with same color).
}\label{fig_sim_harda} }
\end{figure}

To illustrate and verify the analytical results, we performed
numerical simulations. We specify the class border (black-white
dashed line in Fig.~\ref{fig_sim_harda}) composed of two sections.
One section is a segment of the line $a_1+a_2=A$, and the other
one is an arc of the circle $a_1^2+a_2^2=A^2/2$. The segments are
connected at the point $a_1=a_2=A/2$, forming a smooth curve.

The negative class is the bounded part of the first quadrant of
the plane $(a_1, a_2)$, separated by the border. The training
sequence of length $N_{\text{train}}$ (white filled circles in
Fig.~\ref{fig_sim_harda}) is randomly sampled from the negative
class. The positive class is additionally bounded by condition
$a_1^2+a_2^2 < B^2$ with $B>A$.

The master population of the classifier cells is obtained by
randomly sampling the parameters $(m_1, m_2)$ from the log-uniform
distribution in the parameter space, bounded by the minimal and
maximal values $m_{\min}$ and $m_{\max}$. The total number of
cells in the master population is $N_{\text{master}}$. The uniform
density of cells per logarithmic unit of the parameter space is
\beq
\alpha=\frac{N_{\text{master}}}{(\log m_{\max}-\log m_{\min})^2}.
\eeq

The classifier is trained by presenting sequentially all training
samples from the negative class, and discarding all cells
answering positively to at least one sample. Algorithm description
in Table~\ref{alg_hardlearn} formalizes the above procedure.

In our simulation we let $N_{\text{master}}=300$,
$N_{\text{train}}=200$, $A=6$, $B=8$. The smaller border intercept
is $A/\sqrt{2}\approx 4.24$. In accordance to the criterion
formulated in the end of the previous subsection, we let
$m_{\max}=0.5>1/4.24$, and $m_{\min}=m_{\max}/100$. We measure the
quantity of the positively responding cells of the trained
classifier as a function of the input $(a_1, a_2)$. The result is
depicted in Fig.~\ref{fig_sim_harda} in color code. The straight
interfaces of color, distinguishable in the figure, are the
borders of type \eqref{eq_Aborder} associated with the individual
linear classifiers (cells).

\section*{Soft classification problem}
The approach considered above can only be applied to hard
classification problems with a special type of the classification
border (namely, the border must be a curve connecting the axes in
the input space, having a negative slope at each point, with the
negative class being a convex region, see subsection ``Hard
classification principle and learning strategy'' for details). In
order to address problems with classification border of more
general type, or ``soft'' classification problems (i.e. problems
with inseparable classes with a-priori unknown probability
distributions in the input space) we employ soft learning strategy
and a two-input elementary classifier design with a bell-shaped
response function, which was suggested in \cite{FirstPaper}.

\begin{figure}
\centering
\includegraphics[width=0.8\textwidth,clip=true]{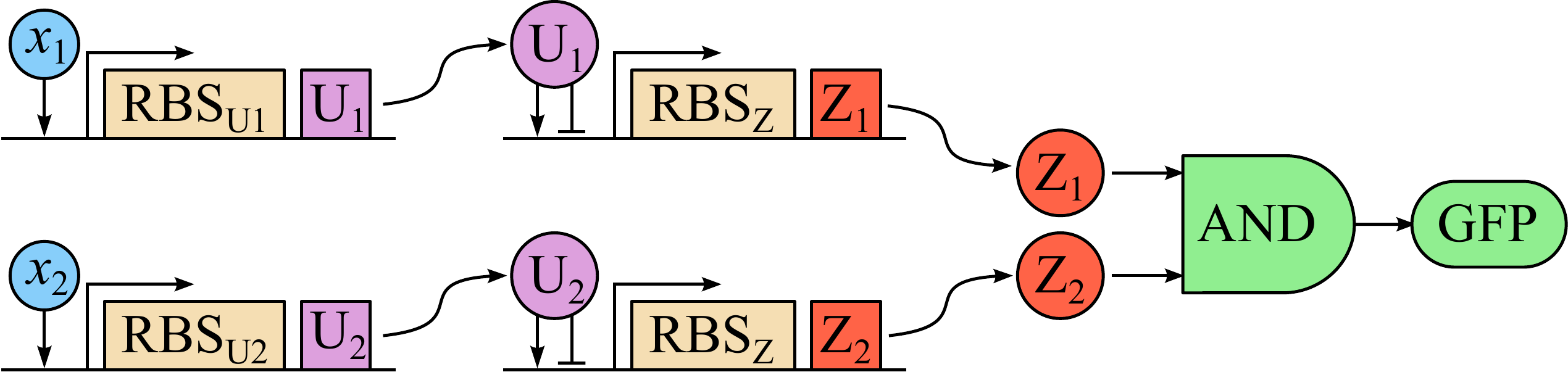}
\caption{{\bfseries Scheme of a two-input classifier circuit with
a bell-shaped response.} $x_1$, $x_2$ -- inputs inducing the
corresponding promoters, $\text{RBS}_{\text{U1}}$ and
$\text{RBS}_{\text{U2}}$ -- ribosome binding sites determining the
strengths of the input branches, $\text{U}_1$, $\text{U}_2$ --
intermediate repressor/activator factors, $\text{Z}_1$,
$\text{Z}_2$ -- outputs of the individual branches, GFP --
reporter gene.}\label{fig_andgate}
\end{figure}

\subsection*{Two-input classifier with a bell-shaped response}
An elementary classifier circuit providing a bell-shaped response
in the two-dimensional input space can be constructed of two
independent sensing branches, whose outputs are combined using a
genetic AND gate (Fig.~\ref{fig_andgate}) \cite{FirstPaper}. Each
sensing branch is composed of two genetic modules, the sensor and
the signal transducer \cite{FirstPaper}. The sensing module is
monotonically induced by the corresponding input chemical signal
$X_j$ ($j=1,2$) and drives the synthesis of an intermediate
repressor/activator $U_j$. The signal transducer part is activated
by $U_j$ at intermediate concentrations and inhibited at higher
concentrations, providing the maximal response at a certain
concentration level. The classic well-characterized example of
such promoter is the promoter $P_{RM}$ of phage lambda which
provides this kind of non-monotonic response to the lambda
repressor protein CI \cite{ptashne1986genetic}.

The outputs $Z_j$ of both sensing branches drive the expression of
a reporter protein (e.g., GFP) through a two-input genetic AND
gate. A number of circuits performing logical operations including
AND have been developed and characterized recently
\cite{wang2011engineering,moon2012genetic,shis2013library}. When
each sensing branch provides a bell-shaped response function, then
the response of the full circuit will also be a bell-shaped
function in the two-dimensional input space.

Omitting the indices $j$ at all variables and parameters for the
sake of conciseness and denoting the concentrations of $X$, $U$,
$Z$ with $x$, $u$ and $z$, the steady-state concentration of each
single sensing branch output $Z$ can be written as
\cite{FirstPaper}
\beq\label{eq_zofx}
z(x; m_u, m_z)=\frac{r_z( r_u(x;m_u)/\mu_u; m_z )}{\mu_z},
\eeq
where $x$ is the input concentration, $\mu_u$ and $\mu_z$ are the
degradation rates of $U$ and $Z$, respectively; $r_u(\cdot)$ and
$r_z(\cdot)$ are the effective production rates of $U$ and $Z$
described by standard Hill functions
\beq
r_u(x;m_u)=m_u\cdot \frac{\alpha A_{u}^{p_u} + x^{p_u}}
{A_{u}^{p_u}+ x^{p_u}},
\eeq
\beq\label{eq_rzofu}
r_z(u;m_z)=m_z\cdot \frac{A_{z}^{p_z} u^{p_z}} {(A_{z}^{p_z}+
u^{p_z})^2},
\eeq
where $\alpha$ determines the basal expression from the sensor
promoter in the absence of the input chemical $X$, $A_u$ and $A_z$
are the dissociation constants of $X$ and $U$ with their
corresponding promoters, the Hill coefficients $p_u$ and $p_z$
characterize the cooperativity of activation or repression of the
corresponding promoters, $m_u$ and $m_z$ describe the overall
expression strength of $U$ and $Z$.

The function $z(x)$ defined by Eqs.
\eqref{eq_zofx}--\eqref{eq_rzofu} is bell-shaped in a range of
$m_u/\mu_u \in (A_z, A_z/\alpha)$, with the position of the
maximum determined by the value of $m_u/\mu_u$ \cite{FirstPaper}.
A master population of elementary two-input classifiers with
response maxima randomly varied in the input space can be
constructed by random variation of the sensory promoter strengths
$m_u$ both among the individual cells, as well as among the two
sensory branches in each cell. The variation range of the maximum
position is limited by the parameter $\alpha$, which is for common
promoters of the order of $10^{-3}$
\cite{lutz1997independent,cox2007programming}. The full range can
be covered, provided the promoter strengths $m_u$ are varied at
least $1/\alpha=10^3$ fold, which is achievable, for example, by
varying the DNA sequence within and near the ribosome binding site
of the sensory gene \cite{salis2009automated,kudla2009coding}.

In the following we let the $m_u$ parameters of the two sensory
branches in a chosen $i$th cell take on the values $m^i_1$ and
$m^i_2$, the lower index denoting the input, the upper being the
cell number, with all other parameters being the same in both
sensory branches in all cells.

We model the AND gate, which drives the reporter protein
production, as a product of two Hill functions
\beq\label{eq_funcg}
g(z_1,z_2)=\beta \cdot \frac{z_1^{p_g}}{A_g^{p_g}+z_1^{p_g}} \cdot
\frac{z_2^{p_g}}{A_g^{p_g}+z_2^{p_g}},
\eeq
where $z_{1,2}$ are the inputs to the AND gate, $\beta$ is a
dimensional constant, $A_g$ and $p_g$ are respectively the
dissociation constant and the Hill coefficient for the AND gate
(for simplicity we assume equal values for both inputs).

The inputs to the AND gate are essentially the outputs of the
sensory branches, thus the output of a chosen $i$th cell finally
is
\beq\label{eq_fiofx}
f_i(x_1,x_2)=g(z(x_1;m^i_1), z(x_2;m^i_2)),
\eeq
where $x_{1,2}$ are the classifier inputs, the function
$g(\cdot,\cdot)$ is defined by \eqref{eq_funcg}, and $z(\cdot)$ by
Eqs. \eqref{eq_zofx}--\eqref{eq_rzofu} with $m_u$ substituted by
$m^i_1$ or $m^i_2$ for either input branch, and index $i$ labeling
the individual cells.

\subsection*{Soft learning strategy}
By ``soft learning'' we mean a learning strategy which reshapes
the population density in the parameter space in response to a
sequence of training examples in order to maximize the correct
answer probability for the distributed classifier taken as a
whole, without any hard separation of the cells into ``correct''
and ``incorrect''.

This can be achieved by organizing a kind of population dynamics
which gives preference to cells which tend to maximize the
performance of the whole classifier. In the simplest case, the
training examples are sequentially presented to all cells in the
population, and some cells get eliminated from the population in a
probabilistic way, with survival probability depending upon the
cell output, given the a-priori knowledge about the particular
training example to belong to a certain class. This may be
implemented by means of FACS technique.

We use a more elaborate learning strategy incorporating a
mechanism for conserving the total cell count. In the model
description this is achieved by simply replacing each discarded
cell with a duplicate of a randomly chosen cell from the
population. In experimental implementation the same effect can
arise from the cell division process which goes on during the FACS
cell selection. Alternatively, the proper competitive population
dynamics can be organized by modulating the viabilities of the
cells in accordance with the learning rule by means of e.g.
antibiotic resistance genes controlled by the classifier circuit
output in each individual cell.

In consistency with \cite{FirstPaper}, we specify the
probabilities of cell survival after presenting each training
example as
\begin{subequations}\label{eq_ppp}
\begin{align}
p_+(g) &= \frac{1}{1+\xi}+\frac{1}{1+\xi
\exp(-g/\gamma)}, \label{eq_pplus} \\
p_-(g) &= \frac{1}{1+\xi}-\frac{1}{1+\xi \exp(-g/\gamma)} +1,
\label{eq_pminus}
\end{align}
\end{subequations}
where $g$ is the cell output upon presenting a training example,
$\xi=\exp((8\gamma^{-1}))$, $\gamma$ controls the ``softness'' of
the learning (the greater $\gamma$, the softer is the slope of
$p_+(g)$ and $p_-(g)$). Either $p_+(g)$ or $p_-(g)$ is used,
depending on the class to which the training example is a-priori
known to belong. The functions specified in (\ref{eq_ppp}a,b) have
maximal slope at $g=1/8$. The cell output range should be scaled
to cover this value by adjusting the constant $\beta$ in
\eqref{eq_funcg}.

The output of a distributed classifier is the sum of all
individual cell outputs:
\beq\label{eq_output}
f(x_1,x_2)=\sum_{i=1}^{N_c} f_i(x_1,x_2),
\eeq
where $f_i(x_1, x_2)$ is defined by \eqref{eq_fiofx}, and $N_c$ is
the total number of cells.

The classification decision is made by comparing the classifier
output to a threshold $\theta$:
\beq
\mbox{decision}=\begin{cases}
 \mbox{``positive''}, & \mbox{if}\, f(x_1,x_2)\ge \theta, \\
 \mbox{``negative''}, & \mbox{if}\, f(x_1,x_2) < \theta,
 \end{cases}
\eeq
where $\theta$ has to be adjusted after the learning to maximize
the correct answer rate of the classifier.

The classification border is actually a level line of $f(x_1,x_2)$
corresponding to the threshold $\theta$. The aim of the soft
learning is thus to reshape the population and select the optimal
value of $\theta$ in a way that the corresponding level line is
the best approximation of the (unknown a-priori) optimal
classification border. The computational criterion of this
optimality is the maximization of the correct answer rate using
the given training examples.

\subsection*{Simulations}

We used algorithm described in Table~\ref{alg_softlearn} to
implement the soft learning strategy. We demonstrate the use of
the soft classification strategy to solve two problems which are
not solvable with hard distributed classifiers described in
section ``Hard classification problem''. The first example has
separable classes which consist of disjoint regions and thus do
not satisfy the requirements of convexity and negative slopes
which were imposed in subsection ``Hard classification principle
and learning strategy''. The positive class is specified as union
of two circles on the $(x_1, x_2)$ plane, one centered at
$x_1=x_2=A$ with radius $R$, and the other centered at $x_1=x_2=B$
with radius $3R$, and the negative class as union of two ellipses,
one centered at $x_1=A$, $x_2=B$ with semiaxes $R$ and $3R$, and
the other centered at $x_1=B$, $x_2=A$ with semiaxes $3R\sqrt{2}$
and $R\sqrt{2}$, where
\beq
R=\frac{1}{32}\left( 1-10^{-1.5} \right),\quad A=10^{-1.5}+2R,
\quad B=10^{-1.5}+8R.
\eeq

\begin{figure}
\centering
\includegraphics[width=0.4\textwidth,clip=true]{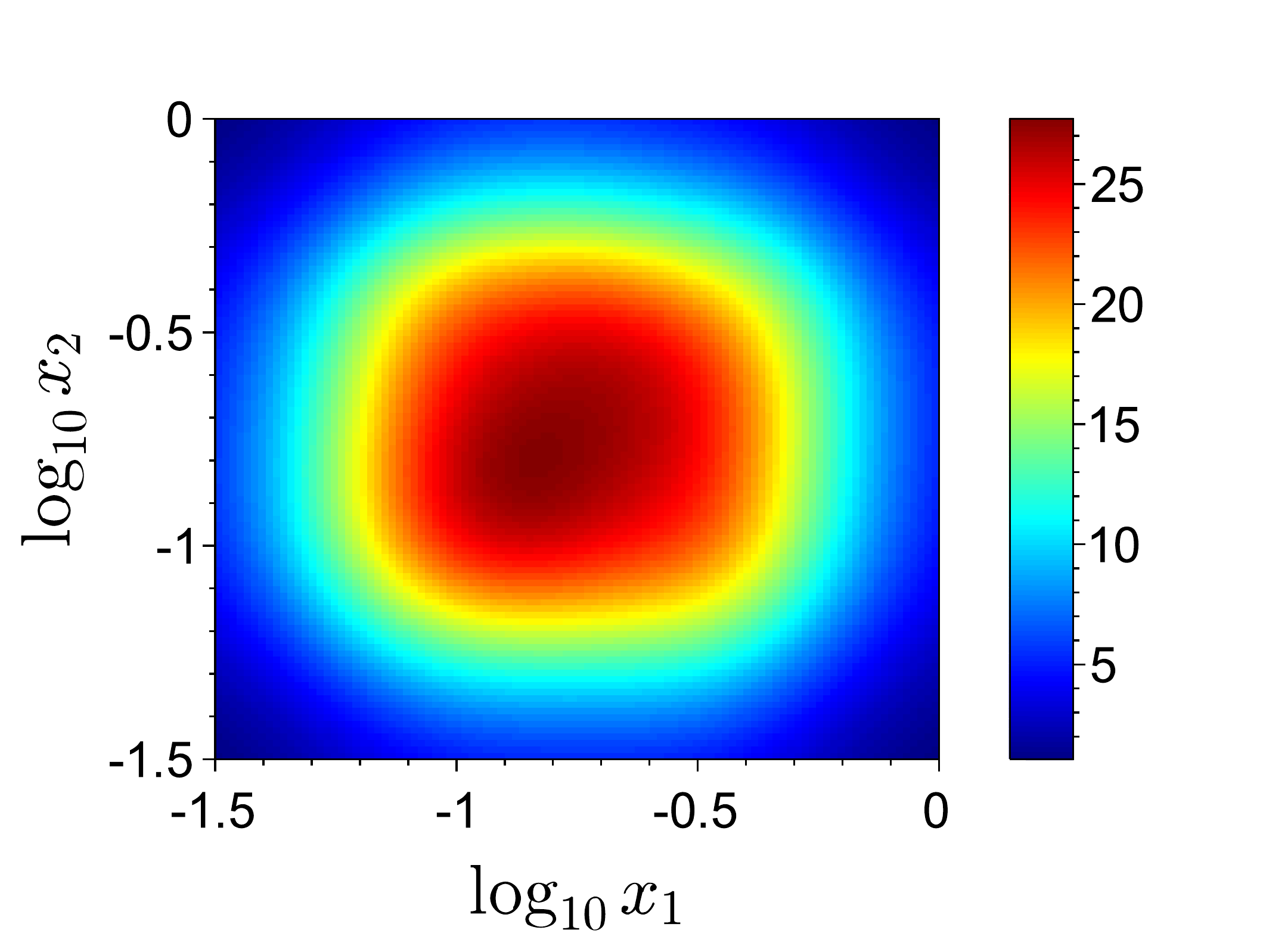}
\includegraphics[width=0.4\textwidth,clip=true]{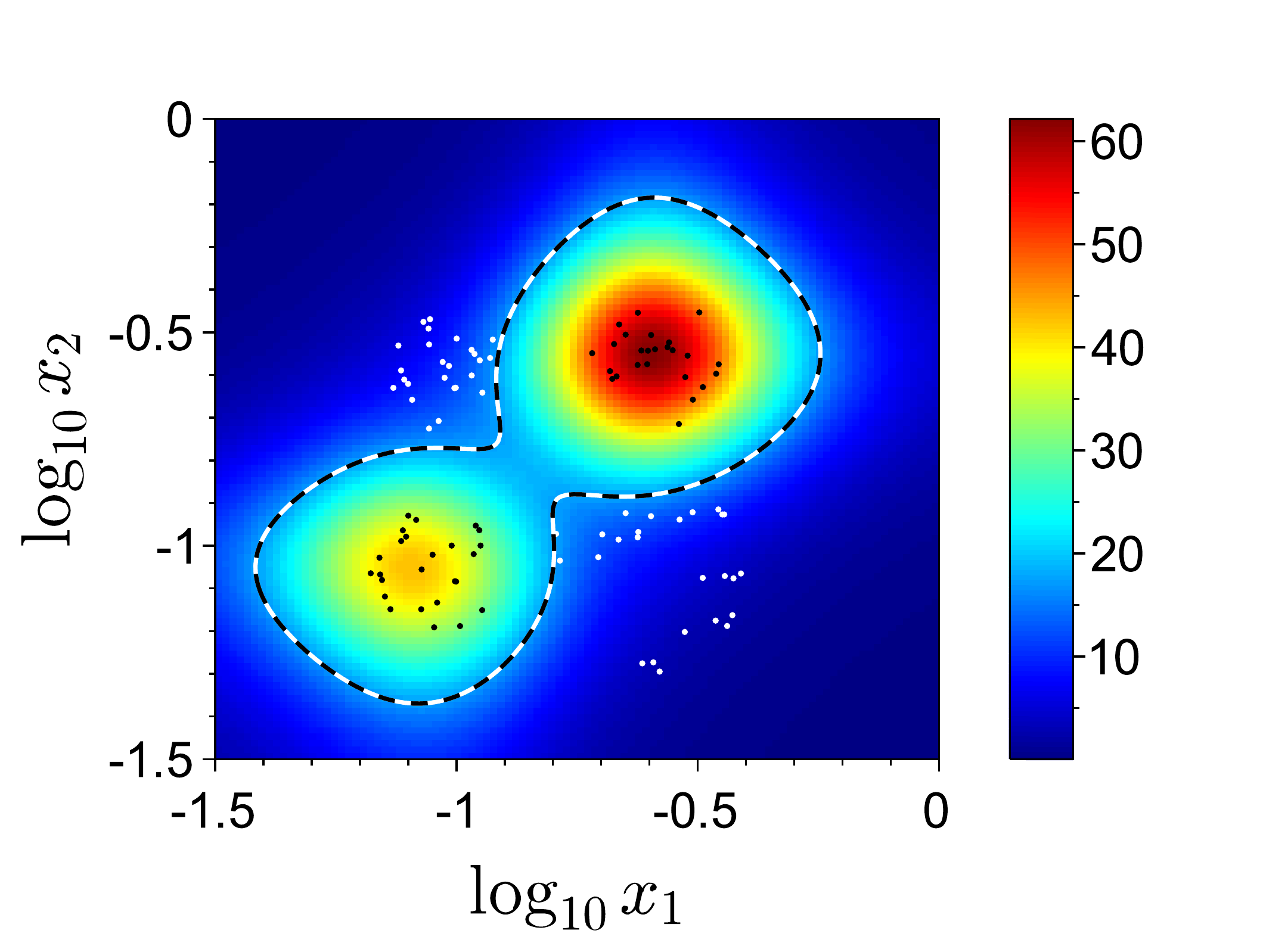}
\caption{{\bfseries Simulation results for soft classification
strategy applied to separable classes.} {\itshape Panel (a).}
Untrained (master) population output (color). {\itshape Panel
(b).} Trained population output (color). White (black) filled
circles -- samples from the negative (positive) class, black-white
dashed line -- classification border of the trained
classifier.}\label{fig_soft1}
\end{figure}

The simulation parameters are $N_c=2\cdot 10^3$,
$N_{\text{train}}=100$ (50 samples from each class),
$N_{\text{iter}}=1000$, softness parameter $\gamma=0.4$,
$m_{\min}=2^2 A_z$, $m_{\max}=2^8 A_z$, $A_z=20$, $m_z=A_u=1$,
$A_g=2$, $p_z=p_u=p_g=2$, $\alpha=10^{-3}$. Output scaling
constant $\beta=1056.25$ is chosen so that cell output $g$ ranges
from 0 to 0.25 in consistency with expressions for survival
probabilities (\ref{eq_ppp}a,b). The simulation result is
presented in Fig.~\ref{fig_soft1}. All training samples are
classified correctly after learning, but this becomes impossible
in case of inseparable classification problems.

\begin{figure}
\centering
\includegraphics[width=0.4\textwidth,clip=true]{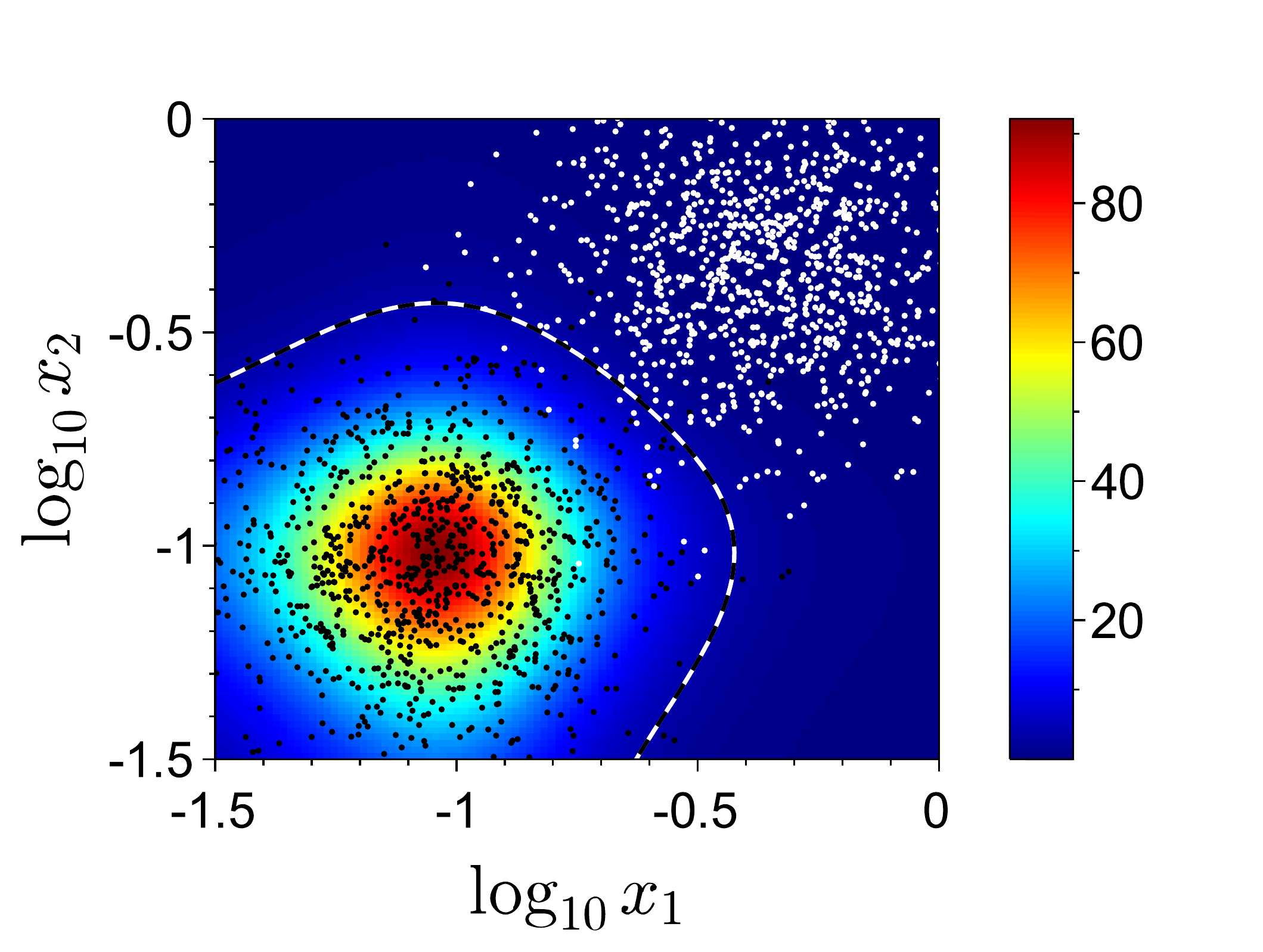}
 \caption{{\bfseries Simulation results for soft classification
strategy applied to inseparable classes.} Notations same as in
Fig.~\ref{fig_soft1}.}\label{fig_soft2}
\end{figure}

The next example shows the classifier operation for inseparable
classes. For either class we use a two-dimensional log-normal
distribution resulting from independently sampling both inputs
$x_1$ and $x_2$ from a one-dimensional log-normal distribution
centered at $\log_{10} x_{1,2}=-1.04$ for the positive class, and
$\log_{10} x_{1,2}=-0.35$ for the negative class, with standard
deviation of $\log_{10} x_{1,2}$ set to $0.22$. The number of
training examples is $N_{\text{train}}=2000$ (1000 samples from
each class). Other simulation parameters are the same as in the
previous example. The result of the simulation is presented in
Fig.~\ref{fig_soft2} the same way as in the previous example.

We compared the successful classification rates of our distributed
gene classifier to that of several common machine learning
algorithms \cite{hastie2009elements} including the $k$-means
method, support vector machine (SVM), and random forest algorithm,
all implementations taken from the ``scikit-learn'' Python library
\cite{pedregosa2011scikit} with default parameters. The comparison
results are presented in Table~\ref{tab_rates}. Simulation~1 is
the same as in Fig.~\ref{fig_soft2}, and the only difference of
Simulation~2 is the log-normal distribution's central point
location for the positive class, namely $\log_{10} x_{1,2}=-0.61$,
which yields a greater overlap of the classes probability
densities. The successful classification rates were computed using
testing sequences of length $N_{\text{test}}=2000$ (1000 samples
from each class), equal to that of the training sequences.

\section*{Discussion}

In summary, in this paper we have presented a design of
multi-input classifiers to be implemented as a synthetic genetic
network. We have considered two examples, corresponding to hard
and soft learning strategy. As a multi input classifier, these
devices can solve classification task based on the data
inseparable in the single dimension case. Moreover, the design
developed allows to achieve practically arbitrary shape of the
classification border in the space of input signals. Here we have
considered two input genetic classifiers but the same design
principles can be utilized to construct multi input classifying
devices, then, the number of inputs is limited only by the number
of possible hybrid promoters.

Our approach challenged a problem of discrimination between classes with
overlapping probability density distributions in the input space.
In this case the classification error probability cannot vanish
and has to be minimized. The optimal solution to this problem is
given by the Bayesian classification rule \cite{duda2012pattern}.
In case of equal a-priori probabilities for a randomly picked
sample to belong to either class, the classification of a
presented sample point from the parameter space is optimally done
by comparing the class probability density functions at this
point: the class with the greatest probability density value is
the optimal answer to the classification problem. At the
classification border the probability density functions get equal.
If these functions are known a-priori, then the optimal border is
thus also known, and the problem reduces to ``hard
classification'' discussed above.

When the probability density functions of the classes are not
known a-priori, the optimal classification rule is not known
either, and the classifier has to be trained by examples. We will
refer to this problem as ``soft classification''. Hard learning is
not applicable in this case, because it may lead eventually even
to discarding all the cells. Inseparable classes with a-priori
unknown probability density functions require another learning
strategy which we will refer to as ``soft learning'', when the
decision to discard or to keep a particular cell upon presenting a
training example is probabilistic, depending on the cell output.

An important aspect of synthetic biology is the design of smart biological devices
or new intelligent drugs, through the development of in vivo digital circuits \cite{2002_Weiss}.
 If living cells can be made to function as computers, one could
envisage, for instance, the development of fully programmable microbial robots
that are able to communicate with each other, with their environment and with
human operators. These devices could then be used, e.g., for detection of hazardous
substances or even to direct the growth of new tissue. In that direction, pioneering
experimental studies have shown the feasibility of programmed pattern formation
\cite{2005_Basu}, the possibility of implementing logical gates and simple
devices within cells  \cite{2012_Moon}, and the construction of new biological devices capable to solve
or compute certain problems  \cite{2009_Haynes}.

The classifiers designed could be considered as a further
development towards the construction of robust and predictable
synthetic genetic biosensors, which have the potential to affect
and effect a lot of applications in the biomedical, therapeutic,
diagnostic, bioremediation, energy-generation and industrial
fields \cite{2009_Lu,2010_Khalil,2011_Ruder,2012_Weber}.

\section*{Acknowledgments}


The research is partly supported by The Ministry of education and
science of the Russian Federation (agreement No.~02.B.49.21.0003).
Authors acknowledge support from Russian Foundation for Basic
Research grants No.~14-02-01202 (RK and AZ) and No.~13-02-00918
(OK and MI), from the Deanship of Scientifc Research, King
Abdulaziz University, Jeddah, grant No.~20/34/Gr (AZ), NIH Grant
RO1-GM069811 (LT), and DARPA Contract W911NF-14-2-0032 (RH).

\bibliography{multi_input,multi_input2}

%
%

\clearpage
\section*{Tables}

\begin{table}[h]
\caption{\bfseries Hard learning algorithm} \label{alg_hardlearn}
\begin{algorithmic}
 \REQUIRE Master population of $N_{\text{master}}$ elementary linear classifiers (cells)
 with parameters $(m_1^i, m_2^i)$ randomly sampled from the log-uniform distribution in the parameter space,
 bounded by the minimal and maximal values $m_{\min}$ and $m_{\max}$. The training sequence of
 negative class samples $(a_1^j, a_2^j)$ of length $N_{\text{train}}$.
 \ENSURE Trained set of cells constituting a distributed
 classifier.
 \FOR{each training sample  $(a_1^j, a_2^j)$}
  \FOR{each cell $i=1$ to $N_{\text{master}}$}
   \IF{\eqref{eq_pos_response} holds for this cell and this input (cell generates a positive
   answer)}
     \STATE Remove the cell from the ensemble.
   \ENDIF
  \ENDFOR
 \ENDFOR
\end{algorithmic}
\end{table}

\begin{table}[h]
\caption{\bfseries Soft learning algorithm} \label{alg_softlearn}
\begin{algorithmic}
 \REQUIRE Master population of $N_c$ elementary classifiers (cells) with bell-shaped output
 with parameters $(m_1^i, m_2^i)$ randomly sampled from the log-uniform distribution in the parameter space,
 bounded by the minimal and maximal values $m_{\min}$ and $m_{\max}$. The sequence of training
 examples $(x_1^j, x_2^j)$ of length
 $N_{\text{train}}$. The known class type $y^j=\pm 1$ for each
 example. The number of training iterations $N_{\text{iter}}$.
 \ENSURE Trained set of $N_c$ cells constituting a distributed
 classifier; classification threshold $\theta_{\text{opt}}$.
 \FOR{iteration $k=1$ to $N_{\text{iter}}$}
  \STATE Choose a random example $(x_1^j, x_2^j)$.
  \FOR{each cell $i=1$ to $N_{c}$}
   \STATE Calculate the $i$th cell output $g_i=f_i(x_1^j, x_2^j)$ according to
   \eqref{eq_fiofx}.
   \STATE Calculate the cell survival probability according to \eqref{eq_pplus} or \eqref{eq_pminus}:
   $p=p_+(g_i)$ if $y_i=+1$, or $p=p_-(g_i)$ if $y_i=-1$.
   \STATE With probability $1-p$, choose a random cell from the population and
   eliminate the $i$th cell, replacing it with the chosen cell.
  \ENDFOR
 \ENDFOR
 \FOR{each training example $j=1$ to $N_{\text{train}}$}
  \STATE Use the trained population to calculate the population output $f(x_1^j,
  x_2^j)$ according to \eqref{eq_output}.
 \ENDFOR
 \STATE Find the optimal classification threshold $\theta_{\text{opt}}$ by maximizing
 the correct classification rate over $\theta$:
 $\theta_{\text{opt}}=\text{argmax} \sum_{j=1}^{N_{\text{train}}}
 y^j \left[ 2H(f(x_1^j, x_2^j)-\theta)-1 \right]$.
\end{algorithmic}
\end{table}

\begin{table}[h]
 \caption{\bfseries Successful classification rate of the distributed
 gene classifier compared to that of other machine learning
 algorithms}\label{tab_rates}
 \centering
 \begin{tabular}{|l||l|l|l||l|}
 \hline
              & $k$-means & Support vector & Random forest & Distributed gene  \\
              &           & machine        &               & classifier        \\
 \hline
 Simulation 1 & 91.3      & 98.9           & 98.3          & 98.35  \\
 \hline
 Simulation 2 & 71.45     & 80.75          & 79.3          & 77.1  \\
 \hline
 \end{tabular}
\end{table}

\appendix
\clearpage
\section*{Appendix S1. Deriving an estimate for hard classifier response}
We assume, that the master population is characterized by a
``working parameter domain'' in the space of parameters $(m_1,
m_2)$, such that the density of cells per logarithmic unit of the
parameter space in this working domain is greater than or equal to
a known minimal value $\alpha$. Precisely, we assume, that the
expected number of cells $dN$ falling within a parameter space
element $dm_1 dm_2$ satisfies the following inequality everywhere
in the working parameter domain:
\beq\label{eq_logunf}
dN \ge \alpha \cdot d(\log m_1)\, d(\log m_2).
\eeq

Below we derive a lower estimate for $N_{pos}(a_1^{\text{in}},
a_2^{\text{in}})$ which is the expectation of the number of cells
answering positively to an input $(a_1^{\text{in}},
a_2^{\text{in}})$ taken from the positive class:
\beq\label{eq_Npos}
N_{pos}(a_1^{\text{in}}, a_2^{\text{in}}) > \frac{2\alpha
\delta^2}{(1+\delta)^2},
\eeq
where $\delta$ is a parameter determining the offset of the input
from the classification border into the positive class. In
general, $\delta$ is defined in a way that $(1+\delta)$ is a
factor by which the negative region has to be scaled so as its
border reaches the input point (see details below). Using polar
coordinates $(\rho,\phi)$ defined by $a_1=\rho \cos \phi$,
$a_2=\rho \sin \phi$, we can define $\delta$ in the form
\beq\label{eq_delta_polar}
\delta=\frac{\rho^{\text{in}}}{\rho_b(\phi^{\text{in}})}-1,
\eeq
where $\rho=\rho_b(\phi)$ is the border equation, and
$(\rho^{\text{in}}, \phi^{\text{in}})$ is the input point:
\beq\label{eq_in_polar}
\begin{split}
a_1^{\text{in}}&=\rho^{\text{in}} \cos \phi^{\text{in}},\\
a_2^{\text{in}}&=\rho^{\text{in}} \sin \phi^{\text{in}}.
\end{split}
\eeq

The derivation of \eqref{eq_Npos} is based upon the assumption
that the whole region of the parameter space associated with
producing the positive output of the trained classifier is
contained in the working parameter domain, and hence is populated
by cells with density satisfying \eqref{eq_logunf}. This is
actually the only requirement limiting the applicability of
\eqref{eq_Npos}.

Below we analyze this requirement in case of the working parameter
domain specified as a rectangle $\{ m_{\min} \le m_{1,2} \le
m_{\max}\}$. We derive a set of conditions providing the
applicability of \eqref{eq_Npos} to a particular input
$(a_1^{\text{in}}, a_2^{\text{in}})$:
\begin{subequations}\label{eq_parcond}
\beq\label{eq_parcond1}
\mu_1^{\ast} \le m_{\max}, \quad \mu_2^{\ast} \le m_{\max},
\eeq
\begin{align}
m_{\min} a_1^{\text{in}} + \mu_2^{\ast} a_2^{\text{in}} &\le 1,\\
\mu_1^{\ast} a_1^{\text{in}} + m_{\min} a_2^{\text{in}} &\le 1.
\end{align}
\end{subequations}
with $\mu_1^{\ast}$ and $\mu_2^{\ast}$ defined as the coefficients
of an equation in the form \eqref{eq_mubord}, describing the
tangent to the border drawn at the point where it is crossed by
the input's radius vector.

Conditions (\ref{eq_parcond}a-c) are given a geometrical
interpretation and can be used for choosing parameter values in
experiment and simulations. Condition \eqref{eq_parcond1} can be
formulated in terms of the border intercepts (the abscissa and the
ordinate of the points where the border crosses the axes). Namely,
$m_{\max}$ should not be less than the inverse of each intercept.
The interpretation of (\ref{eq_parcond}b,c) is less
straightforward, but it suggests that these conditions fail
whenever the input point is too close to either axis, or the
tangent to the border drawn at the point where it is crossed by
the input radius vector is too close to being parallel to either
axis. At the same time, (\ref{eq_parcond}b,c) are the less likely
to fail, the smaller is $m_{\min}$, and the closer is the input to
the border (which implies smaller $\delta$).

We stress that the lower estimate in \eqref{eq_Npos} is obtained
for expectations, and the actual count of the positively answering
cells is a random variate determined by a particular realization
(scattering) of the master population in the parameter space.

\begin{figure}
{\centering
\includegraphics[width=0.4\textwidth]{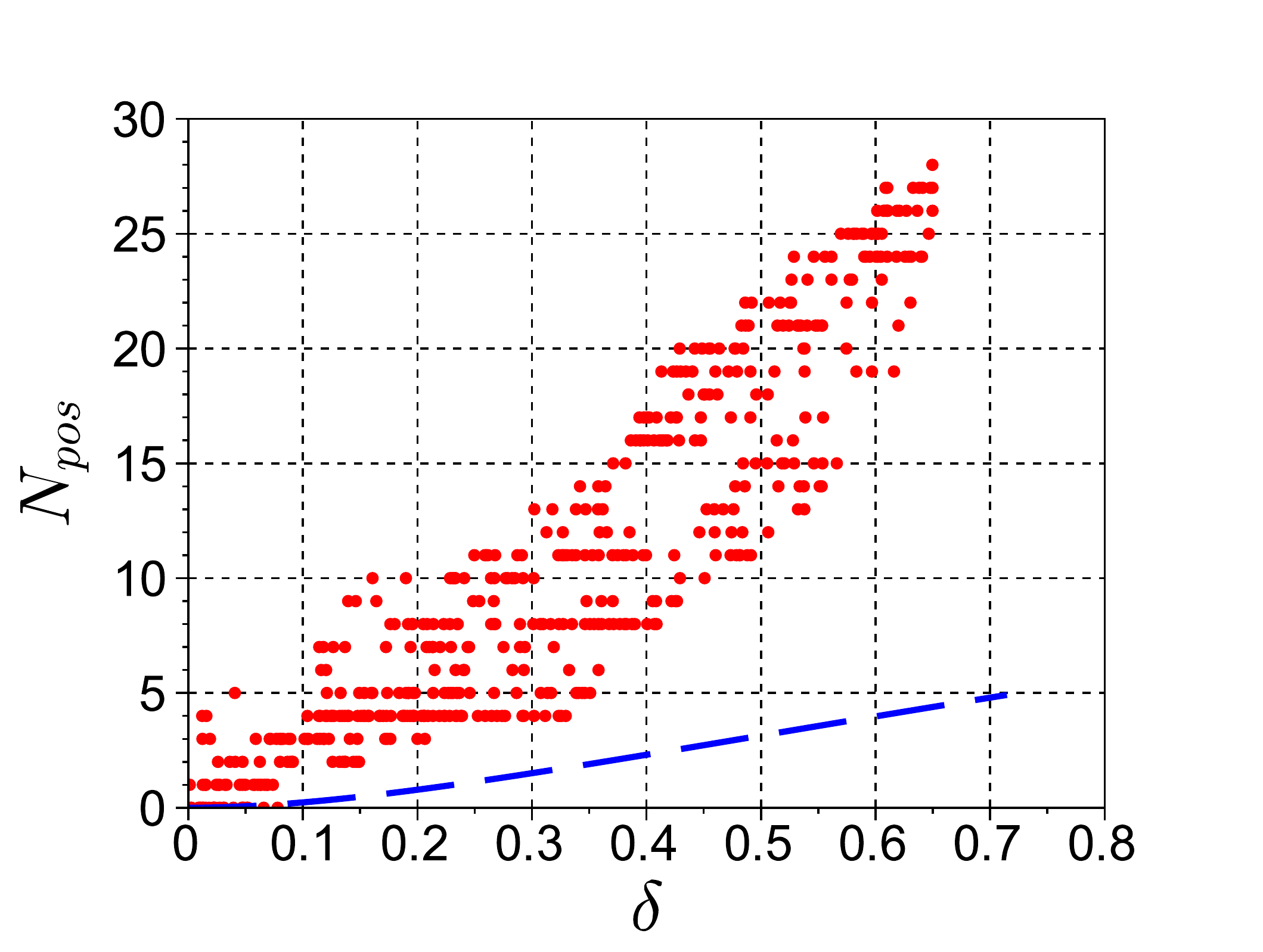}
\caption{{\bfseries Validating the estimate for hard classifier
response.} Number of positively responding cells $N_{pos}$ versus
the input offset $\delta$ from the class border. Red circles --
simulation results, blue dashed line -- lower estimate
\eqref{eq_Npos}. }\label{fig_sim_hardb} }
\end{figure}

To validate the estimate \eqref{eq_Npos}, we extended the
simulation described in section ``Hard classification problem'' by
testing the trained classifier against a sequence of
$N_{\text{test}}=500$ samples from the positive class (black
filled circles in Fig.~\ref{fig_sim_harda}). For each input
$(a_1^{\text{in}}, a_2^{\text{in}})$ we calculate the
corresponding $\delta$ according to \eqref{eq_delta_polar},
\eqref{eq_in_polar}, and measure the quantity of the positively
responding cells $N_{\text{pos}}(a_1^{\text{in}},
a_2^{\text{in}})$. The obtained set of pairs $N_{\text{pos}}$
versus $\delta$ is plotted with red filled circles in
Fig.~\ref{fig_sim_hardb}. The analytical lower estimate
\eqref{eq_Npos} is plotted with a blue dashed line.

\begin{figure}
\centering
(a)\includegraphics[width=0.4\textwidth,clip=true]{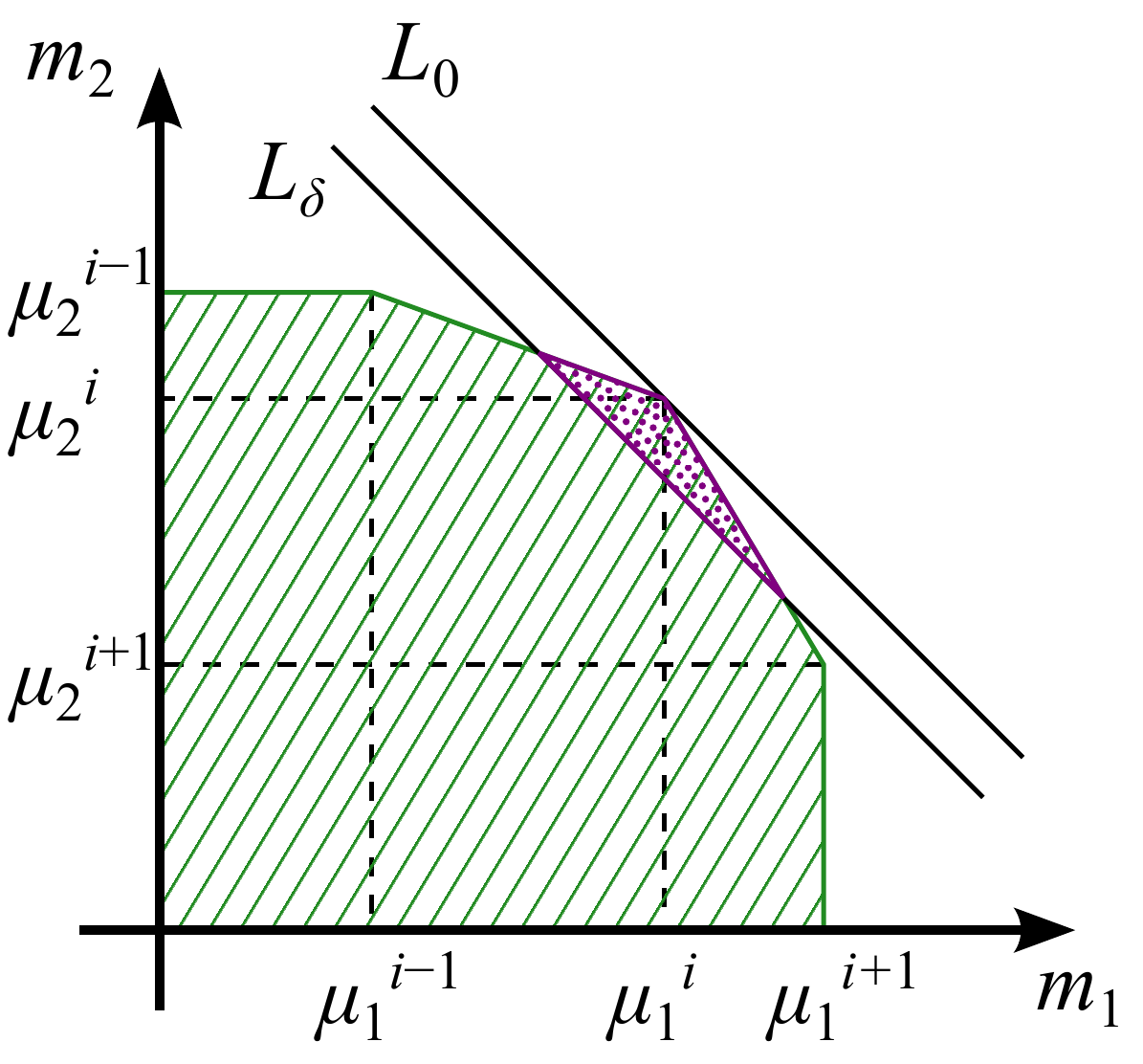}
(b)\includegraphics[width=0.4\textwidth,clip=true]{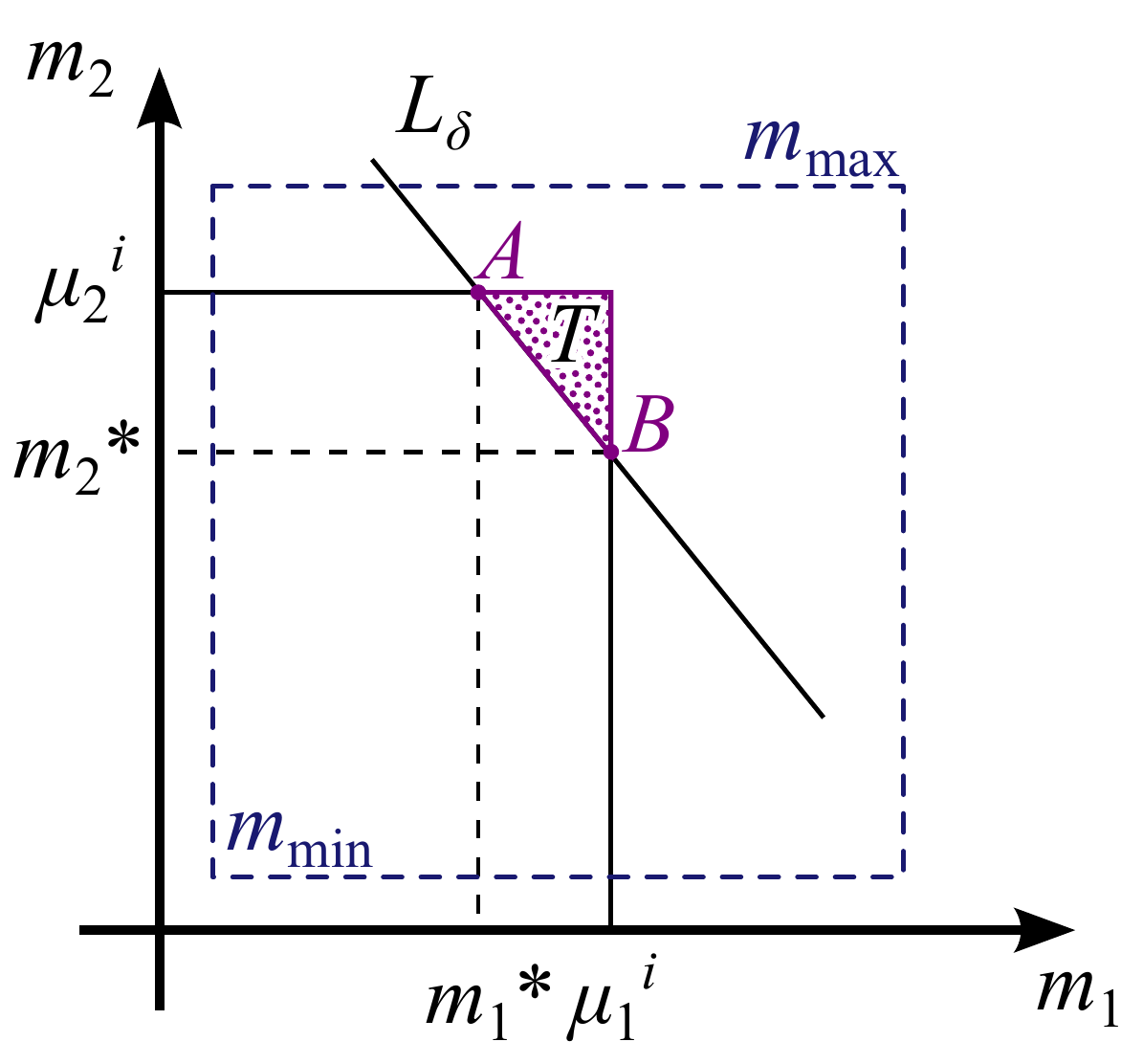}
\caption{{\bfseries Estimating the number of positively responding
cells.} {\itshape Panel (a).} Hatched area -- trained ensemble
region, dotted area -- cells answering positively to an input
sample from the positive class. {\itshape Panel (b).} Dashed
square -- working parameter domain. Other notations -- see text.
}\label{fig_polylearn}
\end{figure}

In order to derive \eqref{eq_Npos}, we first notice, that each
particular input $(\tilde a_1, \tilde a_2)$ can be associated with
a straight line on the parameter plane $(m_1, m_2)$, defined by
the equation
\beq\label{eq_inputline}
m_1 \tilde a_1 + m_2 \tilde a_2 =1.
\eeq

Consider a polygonal classification border (satisfying the
requirements of negative slopes and convexity) and an input
$(a_1^0, a_2^0)$ lying exactly on the border, namely, on its $i$th
segment, and satisfying the $i$th segment's equation:
\beq\label{eq_ithseg}
\mu_1^{i} a_1^0 + \mu_2^{i} a_2^0 =1.
\eeq
The corresponding line $L_0$ defined by \eqref{eq_inputline} with
$\tilde a_1 = a_1^0$, $\tilde a_2 = a_2^0$ then touches the
trained ensemble region on the parameter plane at the vertex
$(\mu_1^{i}, \mu_2^{i})$ (Fig.~\ref{fig_polylearn} (a)).

Now consider another input $(a_1^{\text{in}}, a_2^{\text{in}})$,
which is slightly shifted from the border into the positive class,
namely, satisfying
\beq\label{eq_a12delta}
\mu_1^i a_1^{\text{in}} + \mu_2^i a_2^{\text{in}}=1+\delta,
\eeq
where $\delta>0$ is a parameter determining the offset of the
input point from the negative class. The line $L_{\delta}$ defined
by \eqref{eq_inputline} with $\tilde a_1 = a_1^{\text{in}}$,
$\tilde a_2 = a_2^{\text{in}}$ then crosses the trained ensemble
region (Fig.~\ref{fig_polylearn}~(a)).

Let us estimate the expectation $N_{pos}(a_1^{\text{in}},
a_2^{\text{in}})$ of the number of cells in the trained ensemble
which answer positively to the input $(a_1^{\text{in}},
a_2^{\text{in}})$. We note, that the individual cells answering
positively to this input are exactly those, whose parameters are
located in the trained ensemble region above the line $L_{\delta}$
(dotted area in Fig.~\ref{fig_polylearn} (a)).

We also notice, that the rectangle $\{0<m_1 \le \mu_1^i, 0<m_2 \le
\mu_2^i\}$ is always a subset of the trained ensemble region (see
Fig.~\ref{fig_polylearn} (a)). Denote with $T$ a piece of this
rectangle which is cut from it by the line $L_{\delta}$ (dotted
area in Fig.~\ref{fig_polylearn} (b)):
\beq
T=\left\{ m_1, m_2: 0<m_1 \le \mu_1^i, 0<m_2 \le \mu_2^i, m_1
a_1^{\text{in}} + m_2 a_2^{\text{in}} > 1 \right\}.
\eeq
Denoting the expectation of the cell count in $T$ with
$N_T(a_1^{\text{in}}, a_2^{\text{in}})$, we observe that it is a
lower estimate for $N_{pos}(a_1^{\text{in}}, a_2^{\text{in}})$:
\beq\label{eq_Npos_lowestim}
N_{pos}(a_1^{\text{in}}, a_2^{\text{in}}) \ge N_T(a_1^{\text{in}},
a_2^{\text{in}}).
\eeq
We consider the case when $T$ is a triangle (not a trapezium) and
justify this assumption below.

We express $N_T(a_1^{\text{in}}, a_2^{\text{in}})$ as an integral
\beq\label{eq_integr}
N_{T}(a_1^{\text{in}}, a_2^{\text{in}})=\int_T f(m_1, m_2)\, dm_1
dm_2,
\eeq
where $f(m_1, m_2)$ is the ``cell density function'' in the
parameter space $(m_1, m_2)$. Since the minimal cell density
$\alpha$ in the logarithmic parameter space is specified by
\eqref{eq_logunf}, the cell density function satisfies
\beq\label{eq_logdensity}
f(m_1, m_2) \ge \frac{\alpha}{m_1 m_2},
\eeq
as soon as the pair $(m_1,m_2)$ belongs to the working parameter
domain. We assume, that \eqref{eq_logdensity} holds in the whole
area of $T$ and provide a sufficient condition for this below.

As $f(m_1, m_2)$ in \eqref{eq_logdensity} is falling in both
arguments, the integral in \eqref{eq_integr} can be given a lower
estimate
\beq\label{eq_Nt_lowestim}
N_{T}(a_1^{\text{in}}, a_2^{\text{in}}) > f(\mu_1^i, \mu_2^i)
\cdot S(T) \ge \frac{\alpha \delta^2}{2 \mu_1^i \mu_2^i
a_1^{\text{in}} a_2^{\text{in}}},
\eeq
where $S(T)={\delta^2}/{(2 a_1^{\text{in}} a_2^{\text{in}})}$ is
the area of the triangle $T$. Taking into account the inequality
of arithmetic and geometric means, which along with
\eqref{eq_a12delta} yields
$$\mu_1^i \mu_2^i a_1^{\text{in}}
a_2^{\text{in}} \le (\mu_1^i a_1^{\text{in}}+\mu_2^i
a_2^{\text{in}})^2/4=(1+\delta)^2/4,$$ and combining
\eqref{eq_Nt_lowestim} with \eqref{eq_Npos_lowestim}, we finally
arrive at  \eqref{eq_Npos}.

The input offset parameter $\delta$ is introduced in
\eqref{eq_a12delta}. Geometrically, the input point
$(a_1^{\text{in}}, a_2^{\text{in}})$ is located on a straight line
which results from scaling the line drawn through the $i$th border
segment (defined by \eqref{eq_ithseg}) by a factor of $(1+\delta)$
with the transform center placed at the origin of the coordinates.
This could be used as a definition for $\delta$, but the choice of
the ``$i$th segment'' itself may be ambiguous. However, the
calculation remains valid regardless of this particular choice. To
obtain the best (highest) estimation in \eqref{eq_Npos}, we should
choose the segment number $i$ which maximizes $\delta$ in
\eqref{eq_a12delta} for a given input.

To solve this maximization problem, consider a uniform scaling of
the negative region (with the transform center at the coordinates
origin) by a factor of $s$, such that the input point
$(a_1^{\text{in}}, a_2^{\text{in}})$ becomes located on the scaled
class border (the negative class is ``inflated'' till it reaches
the input point). Denote with $l$ the number of the border
segment, which hits the input point when the border is scaled. It
means, that the following equation is satisfied:
\beq\label{eq_finddelta1}
\frac{\mu_1^l}{s} a_1^{\text{in}} + \frac{\mu_2^l}{s}
a_2^{\text{in}}=1.
\eeq
At the same time, due to the convexity of the negative region, for
all segments of the scaled border the following inequality holds:
\beq\label{eq_finddelta2}
\frac{\mu_1^i}{s} a_1^{\text{in}} + \frac{\mu_2^i}{s}
a_2^{\text{in}}\le 1
\eeq
with the equality taking place only for $i=l$ and, in the special
case when the input hits a vertex of the scaled polygon, for two
adjacent segments. Comparing \eqref{eq_finddelta1} and
\eqref{eq_finddelta2} to \eqref{eq_a12delta} we conclude, that
$\delta$ in \eqref{eq_a12delta} is maximized at $i=l$, and this
maximal value satisfies $1+\delta=s$. Essentially, the ``optimal''
segment $i=l$ is the one which is crossed by the input's radius
vector (or just a straight line segment drawn from the coordinates
origin to the input point).

Thus, for an arbitrary given polygonal classification border
(satisfying the requirements of convexity and negative slopes) the
best estimate in \eqref{eq_Npos} is obtained, when $\delta$ is
defined in a way that $(1+\delta)$ is a factor by which the
negative region has to be ``inflated'' (i.e. scaled up in the
transformed input space $(a_1,a_2)$ with the origin of the
coordinates used as the scaling transform center), so as the input
point finds itself on the scaled classification border. This
definition of $\delta$ remains equally valid in the limit of a
smooth border. Using polar coordinates, we can express this
definition in the form \eqref{eq_delta_polar}.

In the derivation of \eqref{eq_Npos} exactly two assumptions were
made: (i) $T$ being a triangle, and (ii) cell density estimation
\eqref{eq_logdensity} valid in the whole area of $T$. Let us check
their applicability for a given input $(a_1^{\text{in}},
a_2^{\text{in}})$.

In case of a polygonal border, denote with
$\mu_{1,2}^{\ast}=\mu_{1,2}^l$ the equation coefficients of the
``optimal'' border segment $i=l$ (crossed by the input's radius
vector), identified in \eqref{eq_finddelta1}. In the smooth border
limit, instead of the optimal border segment one can speak of the
``optimal'' border tangent, drawn at the point where the border is
crossed by the input's radius vector. In this case we denote with
$\mu_{1,2}^{\ast}$ the equation coefficients of this optimal
tangent.

Denote with $m_1^{\ast}$ the value of $m_1$ at the crossing point
of the lines $L_{\delta}$ and $m_2=\mu_2^{\ast}$ (i.e., abscissa
of the point $A$ in Fig.~\ref{fig_polylearn}(b)), and with
$m_2^{\ast}$ the value of $m_2$ at the crossing point of the lines
$L_{\delta}$ and $m_1=\mu_1^{\ast}$ (i.e., ordinate of $B$ in
Fig.~\ref{fig_polylearn}(b) ):
\beq\label{eq_m_ast}
m_1^{\ast} = \frac{1-\mu_2^{\ast}
a_2^{\text{in}}}{a_1^{\text{in}}}, \quad m_2^{\ast} =
\frac{1-\mu_1^{\ast} a_1^{\text{in}}}{a_2^{\text{in}}}.
\eeq

Assume, that the working parameter domain (where
\eqref{eq_logdensity} holds) is specified as a rectangle $\{
m_{\min} \le m_{1,2} \le m_{\max} \}$. Using the above notations,
we write down the conditions
\begin{alignat}{2}
\mu_1^{\ast} & \le m_{\max}, &\quad \mu_2^{\ast} & \le m_{\max},\label{eq_mastneq1}\\
m_1^{\ast} & \ge m_{\min}, &\quad  m_2^{\ast} & \ge m_{\min},
\label{eq_mastneq2}
\end{alignat}
which provide that $T$ is a subset of the working parameter
domain, and a corollary fact is $T$ being a triangle (since
$m_{1,2}^{\ast}>0$ automatically). Inserting \eqref{eq_m_ast} into
\eqref{eq_mastneq2}, we rewrite it in the form
\beq\label{eq_cond}
\begin{split}
m_{\min} a_1^{\text{in}} + \mu_2^{\ast} a_2^{\text{in}} &\le 1,\\
\mu_1^{\ast} a_1^{\text{in}} + m_{\min} a_2^{\text{in}} &\le 1.
\end{split}
\eeq
Combining \eqref{eq_mastneq1} with \eqref{eq_cond} yields the set
of conditions (\ref{eq_parcond}a-c).

These conditions can be given a geometrical interpretation.
Rewriting the equation of a border tangent in the intercept form
instead of the form \eqref{eq_mubord} yields the intercepts of the
tangent (the abscissa and the ordinate of the points where it
crosses the axes), which are $1/\mu_1^{\ast}$ and
$1/\mu_2^{\ast}$. The minimal values of these intercepts are
actually the intercepts of the border itself (due to convexity).
Therefore, the maximal values of $\mu_1^{\ast}$ and $\mu_2^{\ast}$
can be found as the inverse of the border intercepts. Condition
\eqref{eq_mastneq1} then reduces to the requirement that $m_{max}$
should not be less than the inverse of each intercept.

The interpretation of \eqref{eq_cond} is less straightforward.
Consider the point where the class border is crossed by the
input's radius vector, and the border tangent drawn at this point,
earlier referred to as the ``optimal tangent'', whose equation
coefficients are $\mu_1^{\ast}$ and $\mu_2^{\ast}$. Denote with
$d$ the length of the tangent segment belonging to the first
quadrant and clipped by the axes. This segment is split into two
sections by the tangency point. Denote with $d_1$ and $d_2$ the
lengths of these sections adjacent to the axes $Oa_1$ and $Oa_2$,
respectively, so that $d=d_1+d_2$. Then a geometrical calculation
yields
\beq\label{eq_geometr}
\mu_1^{\ast} a_1^{\text{in}}=(1+\delta)\frac{d_2}{d}, \quad
\mu_2^{\ast} a_2^{\text{in}}=(1+\delta)\frac{d_1}{d}.
\eeq
Inserting \eqref{eq_geometr} into \eqref{eq_cond}, we obtain
\beq\label{eq_geomcond}
\begin{split}
m_{\min} a_1^{\text{in}} + \delta &\le \frac{d_2}{d}(1+\delta),\\
m_{\min} a_2^{\text{in}} + \delta &\le \frac{d_1}{d}(1+\delta).
\end{split}
\eeq

This condition favors smaller $\delta$ and $m_{\min}$, but fails
whenever $d_1/d$ or $d_2/d$ becomes too small. This is the case,
when the input is close to either axis, or the tangent to the
border drawn at the point where it is crossed by the input radius
vector is too close to being parallel to either axis.
\end{document}